\documentclass[doublecol]{epl2}

\title{Scale-free networks resistant to intentional attacks}
\shorttitle{Scale-free networks resistant to intentional attacks}

\author{Lazaros K. Gallos\and Panos Argyrakis}
\shortauthor{L.K. Gallos and P. Argyrakis}
\institute{Department of Physics, University of Thessaloniki, 54124 Thessaloniki, Greece}


\abstract{
We study the detailed mechanism of the failure of scale-free networks under
intentional attacks. Although it is generally accepted that such
networks are very sensitive to targeted attacks, we show that 
for a particular type of structure such networks surprisingly
remain very robust even under removal of a large fraction of their nodes, 
which in some cases can be up to 70\%. 
The degree distribution $P(k)$ of these structures is such that for small values of the 
degree $k$ the distribution is constant with $k$, up to a 
critical value $k_c$, and thereafter it decays with $k$ with the usual power law.
We describe in detail a model for such a scale-free network with this 
modified degree distribution, and we show  both analytically and via 
simulations, that this model can adequately describe all the features
and breakdown characteristics of these attacks.
We have found several experimental networks with such features, such as 
for example the IMDB actors collaboration network or the citations network, whose resilience to attacks
can be accurately  described by our model.
}

\begin{document}

\maketitle

A large number of diverse systems in society, nature and technology can be described
by the concept of a network \cite{AB,DM}. In a network the form of inter-relations between the system
parts determines many structural and dynamic properties of the system. One such property
that has received considerable attention is the robustness of a network under intentional
attack \cite{AJB00,Cohen01}. In the course of such an attack nodes of the network are removed in decreasing
order of their degree $k$ (number of connections to other nodes). This is considered to be the
most harmful type of attack on a network, since the removal of the hubs results in the
largest possible damage. In fact, this vulnerability of the networks to attacks
has been described as their Achilles' heel \cite{AJB00}, because it is generally
accepted that scale-free networks are easily destroyed under intentional attacks. 
This removal process has many and important implications, since
depending on the application, it may describe the resilience of a network, such as the Internet,
or the required number of vaccinations for immunization considerations, etc.

For a scale-free network, where the probability that a node has a given number of links
decays as a power-law, it has been shown that the critical percentage $p_c$ of removed
nodes that results in network disintegration is very low (of the order of a few percent) \cite{Cohen01, Callaway00}. It is,
thus, a well-established  fact, supported by exact analytic results and simulations of
attacks on model and real-life networks, that a scale-free network is very vulnerable
to intentional attacks (where $p_c$ is close to 0), although the same network is
extremely robust under random node failures (where $p_c\simeq 1$) \cite{Cohen00}.

In this Letter we show that there exists a large class of networks that are usually
found in nature and society and have already been characterized as scale-free, but
nevertheless remain robust against removal of the most connected nodes. We first
present the results for real-life networks and then introduce a modified version
for the degree distribution of scale-free networks, for which our analytic and
simulation treatment support these findings.

\begin{figure}
\includegraphics{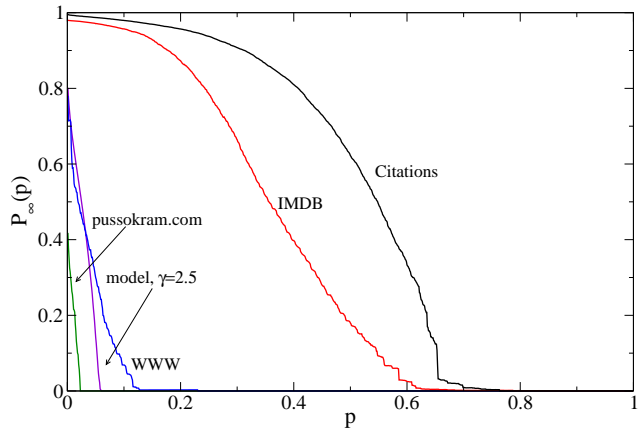}
\caption{\label{fig1}
Percentage of nodes, $P_{\infty}(p)$, belonging to the largest cluster after removal of a fraction $p$ of nodes,
as a function of $p$. The results correspond to intentional attacks on a number of different networks
(shown in the plot).
}
\end{figure}

To demonstrate this issue we performed intentional attacks and random nodes removal to many different
real-life networks.
The critical point was calculated via two distinct methods. In the first method, during the
removal process we monitored the value of the parameter $\kappa\equiv \langle k^2\rangle / \langle k\rangle$,
where $\kappa$ is the connectivity parameter, and
which has been shown to be a measure of the global network connectivity \cite{Cohen00,Paul05}.
A value of $\kappa<2$ signifies the disintegration of a network into isolated clusters.
The second method was a direct measurement of the largest cluster size. The value of $p_c$
was identified as the one where this size assumes for the first time a value close to zero.
The two methods coincide only when the
network is `random' and uncorrelated, in the sense that there is no  inherent organization (or equivalently degree-degree
correlations) in the network.
In a clustered network, though, where these correlations are present, such as the IMDB actors network,
the two methods give different results ($p_c=0.96$ with the first method, but $p_c=0.62$ with the second).
Here, we considered the $p_c$ value derived by the largest cluster size calculation.
The corresponding results for the fraction of nodes $P_{\infty}(p)$ that belong to the largest
cluster of the network during an intentional attack are shown in Fig.~\ref{fig1},
as a function of the percentage $p$ of removed nodes. While the size of the spanning cluster falls 
rapidly in most cases (similarly to a model random network) there are some systems where this
size remains significant even for larger values of $p$.

\begin{table}
\caption{Critical fraction $p_c$ for intentional attacks and random removal on different
networks.}
\label{table1}
\begin{minipage}{0.5\textwidth}
\begin{center}
\begin{tabular}{lcc}
Network & Intentional & Random \\
\hline
Configuration model ($\gamma=2.5$) & 0.055 & 0.99 \\
Online community\footnotemark[1] & 0.04 & 0.90 \\
WWW (nd.edu)\footnotemark[2] & 0.10 & 0.99 \\
IMDB actors collaboration\footnotemark[2] & 0.62 & 0.99 \\
HEP-TH arxiv.org citations\footnotemark[3] & 0.68 & 0.98 \\
\end{tabular}
\footnotetext[1]{(pussokram.com) Data described in Ref.~\cite{Holme04}.}
\footnotetext[2]{http://www.nd.edu/$\sim$networks/resources.htm}
\footnotetext[3]{http://vlado.fmf.uni-lj.si/pub/networks/data/hep-th/hep-th.htm}
\end{center}
\end{minipage}
\end{table}

\begin{figure}
\includegraphics{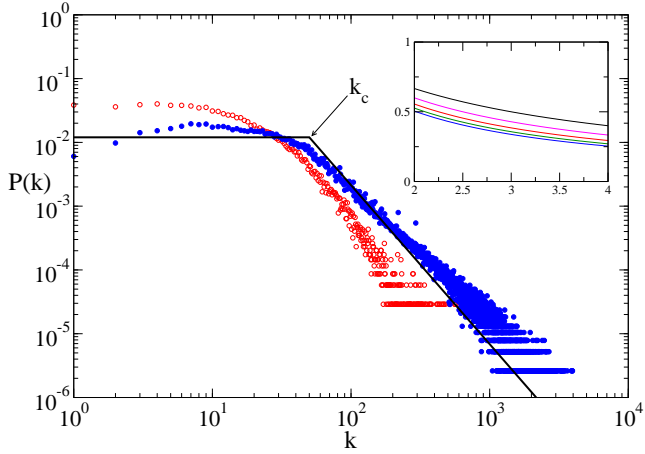}
\caption{\label{fig2}
Degree distributions for IMDB actors (filled symbols) and HEP citations (open symbols).
The solid line represents a typical degree distribution (Eq.~\protect\ref{eq1})
that we used as a model. Inset: Percentage of nodes belonging to the scale-free
part of the distribution as a function of $\gamma$. From top to bottom: $k_c=$2,
3, 5, 10, and 50.
}
\end{figure}

In Table~\ref{table1} we summarize the numerical results we obtained for the critical threshold $p_c$ of the networks
presented in Fig.~\ref{fig1}. Although many of these systems behave in a similar way to the configuration model
network, there is a number of networks, such as actors collaboration and science citations,
where the intentional attack requires removal of a considerable portion of the network nodes,
which is of the order of 65\%.
In order to outline the common feature of these networks, in Fig.~\ref{fig2} we present their degree
distribution. These distributions have a flat or rising part at low-degree
nodes and only after a threshold value the distribution decays as a power-law.
We will show that this feature alone is enough to render a network resistant to attacks,
while the resilience to random node removal remains intact, as we have verified with simulations
that show that in this case the critical threshold remains the same as in simple
scale-free networks, i.e. $p_c\to 1$.

The analytical considerations in the current work apply to simple and random networks,
where connections between nodes are completely random and the network does not include any self-loops
or multiple links between two nodes. The construction of a network for our numerical calculations
follows a slightly modified version of the configuration model. We start with $N$ unconnected nodes
and to each node $i$ we assign a degree $k_i$ from a given distribution $P(k)$, so that each
node has initially a number of unconnected links. We randomly choose two of these unconnected links.
If these links belong to the same node or they belong to two nodes that are already connected we
ignore this selection and randomly choose two other unconnected links. Otherwise, we establish a connection
between these two nodes. We repeat this procedure until all nodes have reached their pre-assigned
connectivity. The use of this method leads to a simple network (i.e. one without self-loops
and multiple links) where the degree distribution follows the pre-defined $P(k)$ function.
We do not impose any upper cutoff for this distribution, so that correlations between degrees
are similar to those of a network with completely random connections and no upper cutoff.

We consider networks whose degree
distribution is uniform for all k values up to a threshold value $k_c$, while for larger k values it decays as a power
law $k^{-\gamma}$, where $\gamma$ is a parameter with typical values in the range 2-4.
The exact form of the distribution, also plotted in Fig. 2 for $k_c=50$ and $\gamma=2.5$, is
\begin{equation}
\label{eq1}
P(k) = \left\{
\begin{array}{ll}
A & 1<k<k_c \\
B k^{-\gamma} & k\geq k_c
\end{array}
\right. \,,
\end{equation}
where the values for the $A$ and $B$ constants are
\begin{equation}
A = \frac{\gamma-1}{k_c\gamma-\gamma+1} \; , \; B=k_c^{\gamma} A \,.
\end{equation}
These values are derived by the requirement that the distribution is properly normalized and continuous.
The fraction of the nodes that belong to the scale-free part of the distribution
(i.e. nodes with $k>k_c$) is shown in the inset of Fig.~\ref{fig2} for different
values of $k_c$ as a function of $\gamma$. We can conclude that the network retains a substantial
scale-free character in practically all cases studied (note also that even for pure scale-free
networks a large portion of the nodes has $k=1$).

We calculate the critical threshold $p_c$ for such a network based on ideas introduced
by Cohen et al \cite{Cohen01} and Dorogovtsev and Mendes \cite{Mendes01}. We employ a continuum approximation
where the degree of a node is treated as a continuous variable.
Nodes are removed according to their initial degree, so that the intentional attack
finally results in the disruption of the network. We consider that the degrees of the nodes
for the network at criticality, i.e. just before disruption, are given by the parameter $\tilde{k}$, with corresponding averages
\begin{equation}
\langle \tilde{k} \rangle = \int_1^{\tilde{K}} k P(k) dk \;,\; \langle \tilde{k}^2 \rangle = \int_1^{\tilde{K}} k^2 P(k) dk \,.
\end{equation}
The effect of an intentional attack is to remove all nodes of a network whose degree is larger
than a cutoff value $\tilde{K}$, i.e. $\tilde{k} \in [1,\tilde{K}]$. This also implies that $p_c$ equals
\begin{equation}
\label{EQresult}
p_c = 1-\int_1^{\tilde{K}} P(k) dk = \int_{\tilde{K}}^\infty P(k) dk \,,
\end{equation}
where the first form is simpler to compute when $\tilde{K}<k_c$ and the second
form when $\tilde{K}>k_c$.
At the same time, removal of a node leads to removing all its links to other nodes. We consider
random networks with no correlations in the nodes connections, which means that a removal of a
node results in removal of random links with probability
\begin{equation}
\label{EQtildep}
\tilde{p} = \frac{\int_{\tilde{K}}^\infty kP(k) dk}{\int_1^\infty kP(k) dk} = 1-\frac{\langle \tilde{k} \rangle}{\langle k \rangle} \,.
\end{equation}

It has been shown \cite{Cohen00,Paul05} that a random network loses its large-scale connectivity after the removal of a critical
fraction $p_c$ of nodes, according to
\begin{equation}
\label{EQpc}
p_c = 1 - \frac{1}{\kappa-1} \,,
\end{equation}
where $\kappa\equiv \langle k^2\rangle / \langle k\rangle$.
This equation has been shown in Ref.~\cite{Cohen01} to be valid for removal of either nodes or links.
As explained in detail there, an intentional attack leads to the equivalent
of a scale-free network with upper cutoff $\tilde{K}$ where a random fraction $\tilde{p}$ of
nodes has been removed. Because of the random character of the network all the links have the
same probability of being removed, and this results to a new degree disribution $\tilde{P}(k)$.
This fact is then used to prove Eq.~(\ref{EQpc}). We can then use this equation for the network resulting after the attack,
by substituting a) $p_c$ with $\tilde{p}$ from Eq.~\ref{EQtildep} and b)
 $\kappa=\langle \tilde{k}^2 \rangle / \langle \tilde{k} \rangle$.
After a few trivial steps Eq.~\ref{EQpc} becomes
\begin{equation}
\label{EQfinal}
\langle \tilde{k}^2 \rangle - \langle \tilde{k} \rangle = \langle k \rangle \,.
\end{equation}
This formula, which is exact, has been already proven in Refs.~\cite{Cohen01,Mendes01}.

In order to use Eq.~\ref{EQfinal} we need to know whether the value of $\tilde{K}$ is
larger or smaller than the threshold value of the distribution $k_c$, so we consider
each case separately.
Calculation of the integrals involved yields
\begin{equation}
\langle \tilde{k} \rangle = \left\{
\begin{array}{ll}
\frac{A}{2}(\tilde{K}^2-1) & \tilde{K}<k_c \\
\frac{A}{2}(k_c^2-1)+ \frac{B}{\gamma-2}(k_c^{2-\gamma}-\tilde{K}^{2-\gamma}) & \tilde{K}>k_c
\end{array}
\right. \,,
\end{equation}
and
\begin{equation}
\langle \tilde{k}^2 \rangle = \left\{
\begin{array}{ll}
\frac{A}{3}(\tilde{K}^3-1) & \tilde{K}<k_c \\
\frac{A}{3}(k_c^3-1)+ \frac{B}{\gamma-3}(k_c^{3-\gamma}-\tilde{K}^{3-\gamma})  & \tilde{K}>k_c
\end{array}
\right. \,.
\end{equation}

The average value of the initial degree distribution $P(k)$ (Eq.~\ref{eq1}) can be approximated with the
assumption that $k_{\rm max}=\infty$. However, for low $\gamma$ values this assumption does not work
well and for a finite-size network we should compute the integral up to the maximum value $k_{\rm max}=K$,
which can be found from
the relation $\int_{k_{\rm max}}^{\infty} P(k) = 1/N $, and is given in our
case by $K= ((\gamma-1)/BN)^{1/(1-\gamma)}$. This results in a correction
to the average value of the unperturbed distribution, which finally becomes
\begin{equation}
\label{EQavk}
\langle k \rangle = \frac{A}{2}(k_c^2-1)+\frac{B}{\gamma-2}k_c^{2-\gamma} - \frac{B}{\gamma-2}\left( \frac{\gamma-1}{BN} \right)^{\frac{\gamma-2}{\gamma-1}} \,.
\end{equation}
The third term is important only for finite-size networks and vanishes as $N\to\infty$.

Combining Eqs.~(\ref{EQfinal})-(\ref{EQavk}) we get
\begin{equation}
\label{EQK}
\begin{array}{ll}
2\tilde{K}^3-3\tilde{K}^2 = \frac{3\gamma k_c^2-4\gamma+8}{\gamma-2} - \frac{6 k_c^\gamma}{\gamma-2} \left( \frac{\gamma-1}{BN} \right)^{\frac{\gamma-2}{\gamma-1}} & \tilde{K}<k_c \\
\frac{k_c}{\gamma-3}\left( \frac{\tilde{K}}{k_c} \right)^{3-\gamma} -
\frac{1}{\gamma-2}\left( \frac{\tilde{K}}{k_c} \right)^{2-\gamma} =  & \\
\frac{\gamma k_c}{3(\gamma-3)} - \frac{\gamma}{\gamma-2} + \frac{2}{3k_c^2} + \frac{k_c^{\gamma-2}}{\gamma-2} \left( \frac{\gamma-1}{BN} \right)^{\frac{\gamma-2}{\gamma-1}}
& \tilde{K}>k_c
\end{array}
\,.
\end{equation}
Solving the above equations for $k_c=\tilde{K}$ we can find the $\gamma$
value for which the lowest degree $\tilde{K}$ of the nodes that need to be removed switches from $\tilde{K}>k_c$ to $\tilde{K}<k_c$. This $\gamma$
value is
\begin{equation}
\label{EQtildeeqkc}
\gamma= \frac{2k_c^3-3k_c^2+4}{k^3-3k^2+2} \,.
\end{equation}

We can now compute the value of $\tilde{K}$ from Eq.~\ref{EQK} and substitute it to Eq.~\ref{EQresult},
which can also be written as
\begin{equation}
\label{EQp}
p_c = \left\{
\begin{array}{ll}
\frac{k_c\gamma-\tilde{K}(\gamma-1)}{k_c\gamma-\gamma+1} & \tilde{K}<k_c \\
\frac{A}{3}(k_c^3-1)+ \frac{B}{\gamma-3}(k_c^{3-\gamma}-\tilde{K}^{3-\gamma}) & \tilde{K}>k_c
\end{array}
\right. \,.
\end{equation}

\begin{figure}
\includegraphics{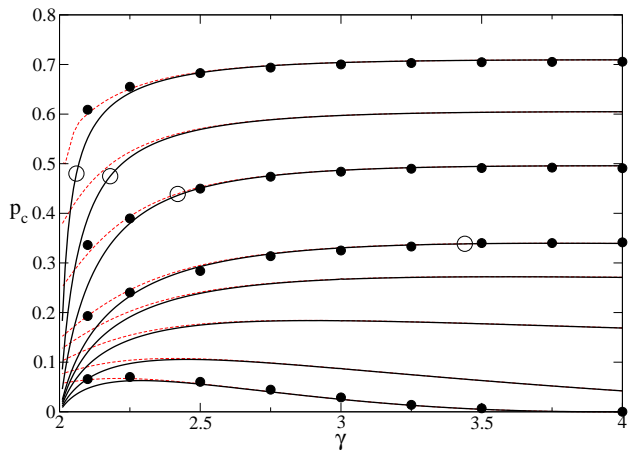}
\caption{\label{fig3}
Critical fraction $p_c$ of removed nodes for networks that undergo an intentional attack,
as a function of the exponent $\gamma$. From top to bottom: $k_c=$ 50, 20, 10, 5, 4, 3, 2 and 1. Thick lines represent the infinite-size numerical solution of Eqs.~\protect\ref{EQK} and \protect\ref{EQp}, dashed lines represent the same solution for $N=10^6$, and filled symbols are simulation results on a network of size $N=10^6$ nodes. The bottom curve for $k_c=1$ is identical to the solution for pure scale-free networks (Ref.~\cite{Cohen01}) The empty
circles denote the solution of Eq.~\protect\ref{EQtildeeqkc}, where the value
of $\tilde{K}$ switches from $\tilde{K}>k_c$ to $\tilde{K}<k_c$.
}
\end{figure}

The numerical solution of Eqs.~\ref{EQK} and \ref{EQp} is shown in Fig.~\ref{fig3} as a function of $\gamma$
for different values of the threshold value $k_c$. In the same figure we also plot results of
simulations on networks that were created with the configuration model. The size of these networks
was $N=10^6$ nodes and their degree distribution obeys Eq.~\ref{eq1}. During the attack process
we removed nodes in decreasing order of their degree and monitored continuously the value of $\kappa$
until it became less than 2. The percentage of the removed nodes up to that point corresponds to the critical value $p_c$.
Note that this method does not have the problems described above, since it is applied to the randomized
networks created via the configuration model. We verified this statement by also comparing to the
results from the largest cluster size method.

Our results for $k_c=1$ coincide with the solution provided in Ref.~\cite{Cohen01}, as can also
be seen numerically from Eqs.~\ref{EQK} and \ref{EQp}.
Comparison of the curves in Fig.~\ref{fig3} for $k_c>1$ to the intentional attack on regular scale-free networks
shows a dramatic increase in the value of $p_c$, over the entire $\gamma$ range. As the threshold
value $k_c$ increases, the stability of the network is further enhanced. Even for $k_c=2$ we observe a significant
influence in the resilience of the network, where $p_c$ is usually more than two times larger than for the case of $k_c=1$.
For $k_c=5$ the critical fraction is already
above 30\%, while when $k_c=10$ the value of $p_c$ lies in the range of
50\%. For even larger values, such as $k_c=50$ which as can be seen in Fig.~\ref{fig1} is not unusual
for real-world networks, the networks exhibit a remarkable resilience to intentional attacks, with a
$p_c$ value close to 70\%. Notice here, that the variation of $p_c$ for $\gamma>2.5$ is almost independent of $\gamma$.
Thus, the important part of the distribution for robustness is the low-degree part and in our model networks
its extent in the $k$-range. On the contrary, an exponent $\gamma>2.5$ for the decaying part does not really
influence the attack result. As the value of $\gamma$ approaches 2, though, the decrease in the value of
$p_c$ is quite sharp, with the infinite-size result $p_c=0$ for $\gamma=2$. For finite size networks
this decrease is much slower and the critical threshold remains significant.

\begin{figure}
\includegraphics{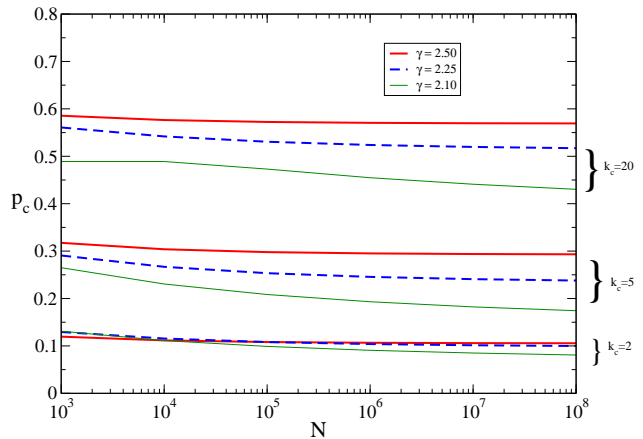}
\caption{\label{fig4}
Variation of the critical threshold $p_c$ with the network size $N$ for different
values of $\gamma$ and $k_c$. The effect of the size on the threshold
is in general not significant, with small exceptions for $\gamma$ values close to 2.
}
\end{figure}

The stability of the solution with respect to the network size $N$ is shown in Fig.~\ref{fig4}.
The value of $p_c$ is practically not influenced by $N$ when $\gamma$ is not close to $\gamma=2$,
such as $\gamma=2.5$ or larger. For these smaller $\gamma$ values the critical threshold exhibits
larger variations, such as in the case of $\gamma=2.1$ presented in the plot. Even in this case,
though, when the network size becomes larger than a moderate size of $N\sim 10^4$ then the critical
threshold remains practically constant.

The explanation behind the enhanced stability can be largely attributed to the increasing average number of connections per node
when the $k_c$ value increases. Although a large value for $\langle k \rangle$ means obviously an enhanced robustness for
the network, we also find that the network is resilient even for very small $k_c$ values. Indeed,
in real networks it is usually not easy to clarify the exact behavior of the degree distribution at very
small degrees, and a difference between $k_c=1$ and $k_c=2$ or 3 can easily be unnoticed.

These findings suggest a structure that is very robust against both random failures and targeted attacks.
This optimization is desirable in most cases and the structure itself, which as we have seen
emerges naturally in many instances, may be used to efficiently protect a network against most attacks.
On the contrary, for immunization purposes, the existence of such networks may present difficulties for
efficient strategies. Even if global knowledge of the entire network structure is available, the required
number of vaccinations remains very high.
In such a case, it is very important to acquire as accurate information on the network structure
as possible, and especially for the low-degree part, because a simple power-law decay of the degree distribution over a large degree range
does not guarantee efficient immunization, if at small values of the degree this power-law decay is
not obeyed.

A study for networks that offer better resilience to attacks than simple scale-free networks
has been performed in Ref.~\cite{Paul04}. The authors find that
the optimal network design for optimization against both random and intentional attack is one
where all nodes have the same degree $k_1$, except for a `central' node with a large degree $k_2\sim N^{2/3}$.
That work, though, has a different scope than ours since the authors kept in all instances the average value $\langle k \rangle$ constant,
while in our work this average value is modified as we modify $k_c$.

In summary, we have studied intentional attacks on networks whose distribution is uniform for low
degrees $k$ and decays as a power law for larger $k$. Such a structure is very robust against both
random and intentional attacks, and outlines the importance of the low-degree nodes in the connectivity
of the structure. Although hubs connect a large part of the network, it is true that they will be unavoidably removed sooner or later,
depending on the removal strategy. However, it seems that the form of the distribution at low
degrees is equally or more important than the existence of the hubs and may render a network vulnerable or stable against intentional
attacks.

\acknowledgments
This work was supported by a NEST/PATHFINDER project DYSONET/012911 of the EC, and also by a project of the Greek
GGET in conjunction with ESF in the frame of international organizations

\end{document}